\documentclass{aastex63}

\usepackage{multirow}
\usepackage{array}
\usepackage{threeparttable}
\usepackage{float}
\usepackage{amsmath}

\received{8 9, 2021}
\revised{9 10, 2021}
\accepted{10 20, 2021}
\submitjournal{ApJS}

\shorttitle{Stellar populations of LAMOST galaxies}
\shortauthors{Li-Li Wang et al.}

\begin{document}

\title{Stellar populations of galaxies in the LAMOST spectral survey}

\correspondingauthor{Li-Li Wang}
\email{jsjxwll@126.com}

\author[0000-0001-7783-9662]{Li-Li Wang}
\affiliation{School of Computer and Information, Dezhou University, Dezhou 253023, China}
\affiliation{Institute for Astronomical Science, Dezhou University, Dezhou 253023, China}

\author{Shi-Yin Shen}
\affiliation{Key Laboratory for Research in Galaxies and Cosmology, Shanghai Astronomical Observatory, Chinese Academy of Sciences, \\
Shanghai 200030, China}

\author{A-Li Luo}
\affiliation{CAS Key Laboratory of Optical Astronomy, National Astronomical Observatories, Beijing 100101, China}
\affiliation{University of Chinese Academy of Sciences, Beijing 100049, China}

\author{Guang-Jun Yang}
\affiliation{School of Computer and Information, Dezhou University, Dezhou 253023, China}

\author{Ning Gai}
\affiliation{School of Physics and Electronic Information, Dezhou University, Dezhou 253023, China}

\author{Yan-Ke Tang}
\affiliation{School of Physics and Electronic Information, Dezhou University, Dezhou 253023, China}

\author{Meng-Xin Wang}
\affiliation{CAS Key Laboratory of Optical Astronomy, National Astronomical Observatories, Beijing 100101, China}

\author{Li Qin}
\affiliation{School of Computer and Information, Dezhou University, Dezhou 253023, China}

\author{Jin-Shu Han}
\affiliation{School of Computer and Information, Dezhou University, Dezhou 253023, China}

\author{Li-Xia Rong}
\affiliation{School of Computer and Information, Dezhou University, Dezhou 253023, China}



\begin{abstract}

We firstly derive the stellar population properties: age and metallicity for $\sim$ 43,000 low redshift galaxies in the seventh data release (DR7) of the Large Sky Area Multi-Object Fiber Spectroscopic Telescope (LAMOST) survey, which have no spectroscopic observations in the Sloan Digital Sky Survey(SDSS). We employ a fitting procedure based on the small-scale features of galaxy spectra so as to avoid possible biases from the uncertain flux calibration of the LAMOST spectroscopy. We show that our algorithm can successfully recover the average age and metallicity of the stellar populations of galaxies down to signal-to-noise$\geq$5 through testing on both mock galaxies and real galaxies comprising LAMOST and their SDSS counterparts. We provide a catalogue of the age and metallicity for $\sim$ 43,000 LAMOST galaxies online. As a demonstration of the scientific application of this catalogue, we present the Holmberg effect on both age and metallicity of a sample of galaxies in galaxy pairs.

\end{abstract}



\section{Introduction}
\label{sec:intro}

Galaxies are systems of stars, gas and dust. A galaxy spectrum in optical wavelength, which is mainly composed of the accumulated light of hundreds of billions of stars, encodes the age and metallicity distributions of stars. From stellar population analysis of galaxy spectra, the star formation and chemical evolution history of galaxies are expected to be decoded. The stellar population synthesis is the most widely used method to estimate the stellar population parameters by comparing the observed galaxy spectra to model spectra\citep{Bica1988, Vazdekis1996, Boisson2000, Bruzual2003, Cappellari2004, cid2005, Ocvirk2006, Tojeiro2007, Maraston2011, Chen2012, Wilkinson2017}. There are many different synthesis algorithms using different characteristic spectroscopic information, e.g. the integrated spectral energy distribution(SED) \citep{Cappellari2004, cid2005, Ocvirk2006, Tojeiro2007, Cappellari2017, Wilkinson2017} and specific absorption line features \citep{Worthey1994, Thomas2003, Kauffmann2003a, Kauffmann2003b, Vazdekis2005, Thomas2011}.

In recent years, a number of full spectral fitting algorithms have been publicly available, such as pPXF\citep{Cappellari2004, Cappellari2017}, STARLIGHT\citep{cid2005}, and FIREFLY\citep{Wilkinson2015, Wilkinson2017}. These methods incorporate the full spectral information in the fitting process, and typically employ a chi--squared minimization approach to calculate the likelihood (or a posterior probability distribution) of galaxy physical properties by comparing observational spectra to a set of linear combinations of single stellar populations(SSPs). The outputs are a best-fitting combination of templates and a set of weights of these templates. Since the full spectral fitting methods use information distributed along all wavelength of the spectrum, the fitting results are sensitive to the flux calibration of galaxy spectra. Although the global shape of the continuum might be compensated by a multiplicative polynominal during the fitting process. However, it has not been systematically tested that how well this multiplicative polynominal could compensate for the inaccuracy of the flux calibration. In practice, the order of the polynomial also can not be well-defined unless we have a detailed understanding of the uncertainties of the flux calibration.

Even if the flux calibration of spectra is good enough, there is still age--dust degeneracy in the full spectrum fitting, especially for the galaxies with young stellar population \citep{Pforr2012, Conroy2013}. To break this degeneracy, new algorithms have been proposed, in which the basic idea is to decompose the small-scale and large-scale signals in the wavelengths. For example, in the full-spectrum fitting code FIREFLY\citep{Wilkinson2015, Wilkinson2017}, a high-pass filter(HPF) is convolved with a observed spectrum so as to remove the long-wavelength mode of the spectrum and then do the spectral fitting. However, when the stellar populations of galaxies are complex, the filtered spectra can not be directly combined from the filtered SSPs(see Appendix \ref{sec:model_decomp} for a detailed discussion). \citet{Li2020} separates the small-scale features from the large-scale spectral shape by performing a moving average method and then fits the observed ratio of the small- to large-scale components(S/L) with the S/L ratios of the SSP models simultaneously. In this method, by fitting the S/L ratios, the derived dust attenuation curves of galaxies could be equally recovered without the knowledge of the continuum shape. In this study, we aim to take the advantages of the fitting method of \citet{Li2020} to estimate the stellar populations of galaxies in the LAMOST spectral survey, whose continuum shapes of galaxy spectra are not very accurately calibrated.

The Large Sky Area Multi-Object Fiber Spectroscopic Telescope (LAMOST) is characterized by its capability of both a large field of view and large aperture with an effective aperture of 3.6--4.9m and 4,000 fibers mounted on its focal plane\citep{Cui2012,Zhao2012}. The wavelength range of LAMOST spectra covers 3700-9000 \AA \ that is recorded in two arms, a blue arm(3700--5900\AA) and a red arm(5700--9000\AA), with a resolution of 1800\citep{Luo2015}. When the blue and red channels are combined together, each spectrum is re--binned in constant--velocity pixels, with a pixel scale of 69 km s$^{-1}$. The seventh data release (DR7) of LAMOST released 10,640,255 spectra, which consists of 9,881,260 star spectra, 198,393 galaxy spectra, 66,406 quasar spectra, and 494,196 unknown object spectra. Although the LAMOST spectral survey focuses on the stellar objects inside Milky Way, there are also some specific targets on extragalactic objects, e.g. QSO identification\citep{Ai2016,YaoS2019}, complementary galaxies to SDSS main galaxy sample\citep{Shen2016, Feng2019}, Complete Spectroscopic Survey of Pointing Area(LCSSPA:\citealt{Wu2016, Yang2018}), bright-infrared galaxies targets selected from infrared surveys \citep{Luo2015}, etc. Besides, some SDSS galaxies have been observed with free fibers. With these LAMOST spectra, many extragalactic objects with peculiar spectroscopic features have been identified \citep{Yanghf2015, YangQ2018, Wangmx2019}. In addition, the nebular emission lines \citep{Wang2018MNRAS} and central velocity dispersion \citep{Napolitano2020} of the galaxies in the LAMOST spectral survey have been measured and published on-line. However, the studies on the stellar population of the galaxies in the LAMOST spectral survey are absent so far. The main difficulty comes from the fact that the flux of LAMOST spectra is not well calibrated\citep{Luo2015}. The LAMOST survey is designed as a spectroscopic survey without a corresponding photometry component. As a result, LAMOST spectra are not calibrated for absolute fluxes, but only for relative fluxes using some standard stars. Because of the uncertainties of galactic reddening on selected standard stars, there are also resulted uncertainties on the global shapes of the continua of LAMOST spectra. In our previous study\citep{Wang2018MNRAS}, we have checked the uncertainties of the flux calibration of LAMOST spectra by comparing synthetic magnitudes of LAMOST spectra with SDSS photometric magnitudes in $ g $, $ r $, $ i $ bands\citep[see][Figure 3]{Wang2018MNRAS}. The median(Quartiles) color difference in high latitude($ |b|\geq60^{\circ} $) is found to be $\Delta (g-r) \approx$ -0.1(-0.21, 0.01) and $\Delta (r-i) \approx$ -0.08(-0.13, -0.03), and in low latitude($ |b|<30^{\circ} $) the median(Quartiles) $\Delta (g-r) \approx$ -0.16(-0.29, -0.03) and $\Delta (r-i) \approx$ -0.11(-0.17, -0.05). We find that the LAMOST spectra are systematically bluer than the SDSS photometry, and the difference is greater as the Galactic latitude is lower.

In this study, we aim to eliminate the effect of the continuum shape in the stellar population synthesis by comparing the small-scale components of the observed spectra with the small-scale components of the SSP models, and then provide the stellar population estimations for the galaxies only spectroscopically observed by the LAMOST spectral survey but not by SDSS. The structure of this paper is as follows. Section 2 describes the galaxy sample in the LAMOST spectra survey. Section 3 presents the details of our spectral fitting algorithm. In Section 4, we make tests on our fitting algorithm using mock spectra and the repeated observations of LAMOST and SDSS galaxies. In Section 5, we apply our method to the galaxy spectra in the LAMOST spectral survey till data release 7(DR7) and present a catalogue of their stellar population parameters. Moreover, we present a scientific application of our catalogue by showing the Holmberg effects of galaxies in galaxy pairs. The summary is given in Section 6. Throughout this study, we adopt the cosmological parameters with $ H_0 $=100 km s$^{-1}$ Mpc$^{-1}$, $ \Omega_M $=0.3, $ \Omega_{\Lambda} $=0.7.

\section{Galaxies in LAMOST DR7}
\label{sec:galLAMOST}

Our study is performed on the DR7 of the LAMOST spectral survey, where 198,393 spectra of 166,691 objects have been identified as galaxies. Among them, about 65\% have spectral counterparts in the sixteenth Data Release(DR16) of Sloan Digital Sky Survey(SDSS)\citep{York2000,ahumada2019sixteenth}. In other words, about 66,648 spectra of 57,581 targets are only observed spectroscopically by LAMOST survey. There are no spectroscopic observations of these targets in SDSS DR16. The target selection of these galaxies in the LAMOST spectral survey includes some specific science aims, for example, the complementary galaxy sample \citep{Shen2016, Feng2019} and the LAMOST Complete Spectroscopic Survey of Pointing Area\citep{Wu2016,Yang2018}, see \citet{Luo2015} for details.

\begin{figure}
	\centering
	\includegraphics[width=8.5cm]{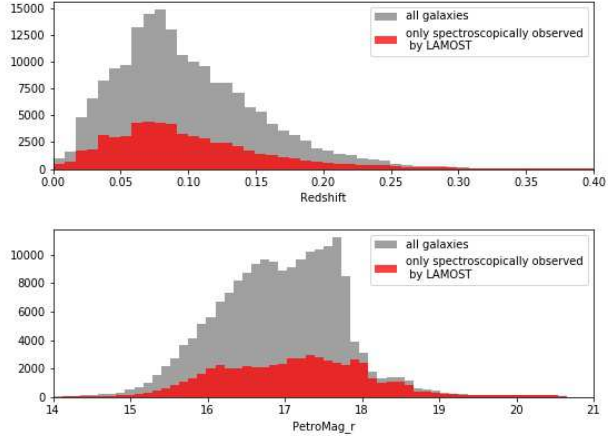}
    \caption{Redshift and magnitude distributions of galaxies in the LAMOST spectral survey DR7. The top panel is the redshift distribution and the bottom is the $r$ band Petrosian magnitude distribution. The grey color represents the distribution of all LAMOST galaxies, while the red labels those of galaxies only spectroscopically observed by LAMOST. }
    \label{fig:mag_z}
   \end{figure}

We show the redshift and $r$ band magnitude distributions of galaxies in the LAMOST spectral survey DR7 in Figure \ref{fig:mag_z}, where the grey color represents the distribution of all galaxies and the red color shows the one of new observations. The $r$ band magnitude is matched from the Petrosian magnitude(petroMag\_$r$) of SDSS DR16 photometric catalogue.\footnote{Most of LAMOST galaxy targets are selected within the SDSS photometric footprint and therefore have SDSS magnitudes. There are a small fraction (5\% ) of galaxy targets without photometric SDSS counterparts (e.g. the bright infrared galaxies observed by LAMOST survey).} The mean $r$ band Petrosian magnitudes for all LAMOST galaxies and only spectroscopically observed by LAMOST are 16.95 and 17.15, respectively, and the mean redshifts of these two samples are 0.998 and 0.103, respectively. As can be seen, both of them are comparable to those of the SDSS main sample galaxies\citep{Stoughton2002}.

In this paper, we focus on the spectra of galaxies which are only spectroscopically observed by LAMOST survey and measure their stellar population parameters. In order to obtain reliable measurements, we restrict the LAMOST spectra with signal to noise ratio in $r$ band(hereafter $S/N_{r}$) $\geq5$. We get $\sim$ 52,000 spectra($\sim$ 43,600 targets) as our measurement data.

\section{Fitting method}
\label{sec:fittingM}

In the typical stellar population synthesis, an observed galaxy spectrum is fitted with a combination of SSPs with different ages and metallicities. There are several widely used SSP models in the literatures \citep{Bruzual2003, Vazdekis2010, Maraston2011, Vazdekis2016, Conroy2018, Maraston2020}. In this study, we use the SSP models from \citet{Vazdekis2010}( hereafter V10) based on empirical stellar library MILES \citep{SanchezBlazquez2006}, which has a wavelength range $\lambda\lambda$3540--7410\AA \ and a spectral resolution of 2.5\AA \ (FWHM)\citep{Falcon-Barroso2011, Beifiori2011}. V10 provides a sample of SSPs covering large stellar parameter regimes: age from 0.06 to 18Gyr and metallicity from -1.71 to 0.22 (log(Z/Z$\sun$)).

As we have mentioned, to avoid the bias from the continua shape, we estimate the stellar population properties by fitting the small-scale features of observed spectrum with the small-scale features of the models. To do that, we first decompose the observed spectra and SSPs into two components: small-scale and large-scale components. The processes of decomposition are detailed as follows.

We calculate the large-scale component of an observed spectrum by average filtering similar to \citet{Li2020}. The window size for filtering is important to get the large-scale component. We have tested window sizes ranging from 100 to 700\AA , and finally choose 300\AA \ as the sliding window size. The testing process is described in details in the Appendix \ref{sec:winsize}. Note that before calculating the large-scale component, several well-known gas emission line regions such as Balmer series and some forbidden-line doublets have been masked.

After obtaining the large-scale component $O_L(\lambda)$, we divide the observed spectrum by its large-scale component to get its small-scale component $O_S(\lambda)$ by Equation \ref{eq:Osmall}. This definition of small component can avoid to be affected by uncertain flux calibration or dust attenuation.\footnote{Although \citet{Li2020} defines $O_S(\lambda)$ = $O(\lambda) - O_L(\lambda)$, they fit the spectra features of $R_{\lambda}=O_S(\lambda)/O_L(\lambda)$ for stellar populations. We can easily find that $R_{\lambda}=O(\lambda)/O_L(\lambda) - 1$. That is to say, our fitting algorithm is similar to theirs.}

\begin{equation}\label{eq:Osmall}
    O_S(\lambda)=\frac{O(\lambda)}{O_L(\lambda)}
\end{equation}

In the conventional full spectral fitting, the model spectrum $M(\lambda)$ is a linear combination of full spectra of SSPs. In order to fit the observed small-scale component, we also separate the small-scale component of $M(\lambda)$ divided by its large-scale component $M_{L}(\lambda)$. The small-scale feature of model spectrum $M_{S}(\lambda)$ is parameterized by,

\begin{equation}\label{eq:small}
   M_{S}(\lambda)= \frac{M(\lambda)}{M_{L}(\lambda)}
=\frac{\sum\limits_{i=1}^{N} a_{i}SSP^{i}(\lambda)}{\sum\limits_{i=1}^{N} a_{i}SSP_L^i(\lambda)}
\end{equation}

From Equation \ref{eq:small}, we can see that the small-scale component of model is a linear combination of full spectra of SSPs divided by a linear combination of large-scale components of SSPs with the same weights $a_{i}$.

After spectral decomposition of observed spectrum, we then make the classical minimum $\chi^2$ fitting of the observed small-scale features with the small-scale component of model:
\begin{equation}\label{eq:chi2}
   \chi^2 = \Sigma_{\lambda} \Bigg [O_{S}(\lambda) - \frac{\sum\limits_{i=1}^{N} a_{i}SSP^{i}(\lambda)}{\sum\limits_{i=1}^{N} a_{i}SSP_L^i(\lambda)}\Bigg ]^2
\end{equation}

\begin{figure}
\centering
	\includegraphics[width=9cm]{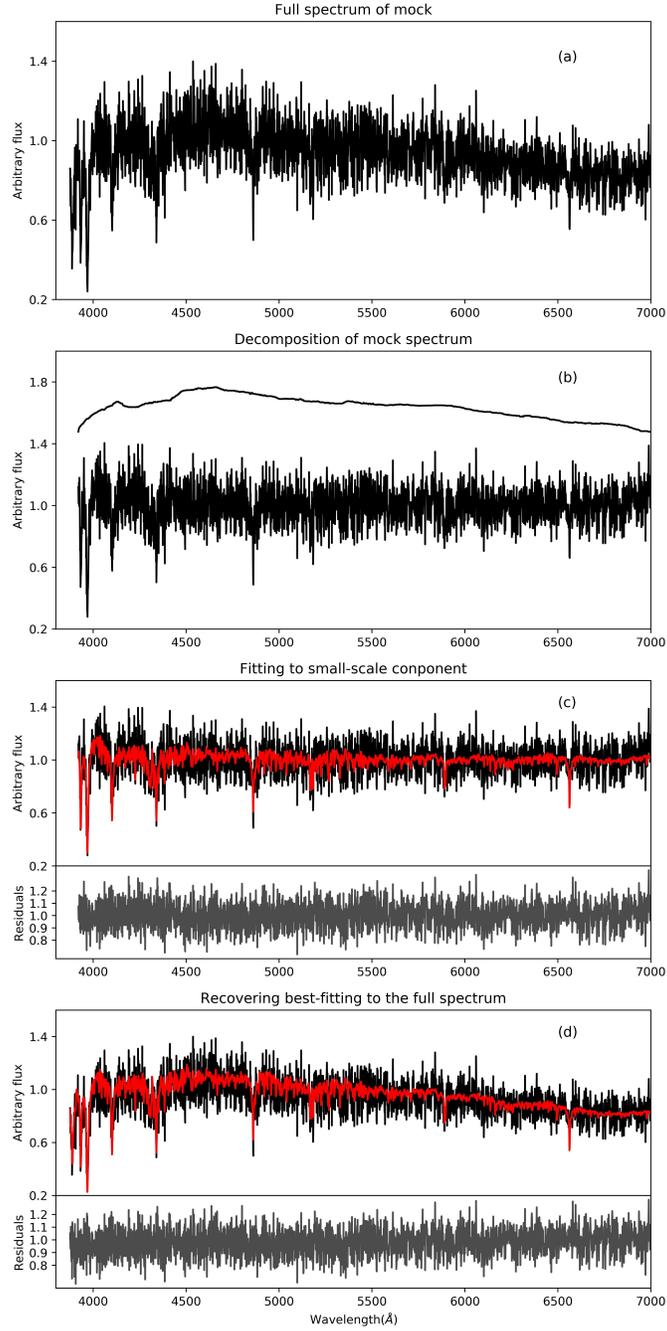}
    \caption{An example of fitting process based on the small-scale features using a mock spectrum. Panel (a) shows the input mock spectrum from V10 SSPs \citep{Vazdekis2010} with $S/N$=10. Panel (b) gives the decomposition of the mock spectrum: the upper curve is the large-scale component and the other is small-scale features. Panel (c) illustrates the spectral fit based on Equation \ref{eq:chi2}, in which the black spectrum is the small-scale features of mock and the red is the best-fitting solution. In the panel (d), the red spectrum is a model spectrum reconstructed from a linear combination of SSPs with the coefficients $a_i$ of our small-scale feature best-fitting, and the black is the full spectrum of mock. In the bottom of panel (c) and (d), we show the residuals of the black spectrum to the red one.}
    \label{fig:fitting}
   \end{figure}

The fitting is performed by using a non-linear least squares minimization routine \emph{LMFIT}\citep{Newville2014}. In order to speed up the fitting process, we select 36 SSPs with 9 ages(0.06, 0.12, 0.25, 0.5, 1.0, 2.0, 4.0, 8.0, 15Gyr) and 4 metallicities (-1.71, -0.71, 0, 0.22) from V10. The ages of SSPs are chosen with approximately equal intervals in logarithm space. For a better fit of the model, we linearly interpolate the SSPs to the same velocity scale of the observed spectra. During the fitting procedure, a set of fractional weights {$a_i$} with minimum $\chi^{2}$ values is determined. And then the best-fitting model spectrum and mean age and metallicity can be obtained. Figure \ref{fig:fitting} illustrates an example of the whole fitting process using a mock spectrum.

(i) In panel (a) of Figure \ref{fig:fitting}, we present a mock spectrum created from V10 SSPs with Gaussian signal-to-noise ratio = 10. The spectrum is the full spectrum with large- and small-scale components. The construction of mock spectra is detailed in next section.

(ii)We apply the average filtering method to trace the large-scale component with a sliding 300-pixel window. The small-scale component is measured by dividing the input spectrum by its large-scale component. The large- and small-scale components are separately plotted in the panel (b) in Figure \ref{fig:fitting}.

(iii)In the fitting, the small-scale component of input spectrum is compared with the one of model spectrum based on Equation \ref{eq:chi2} to determine the coefficients $a_i$ of the best-fitting. The fit result is shown in the panel (c) of Figure \ref{fig:fitting}, where the black spectrum is the small-scale component of mock and the red is the best-fitting small-scale component. We can see from the residuals of input small-scale component to the best-fitting that the fitting result of small-scale components is well.

We reconstruct a full model spectrum by a linear combination of SSPs with the coefficients $a_i$ of the small-scale feature best-fitting. The reconstructed model spectrum includes both large- and small-scale components, which is shown in red in the panel (d). The black spectrum in panel (d) is the full spectrum of mock. From residuals in bottom panel of (d), we can see that the reconstructed full model spectrum is close to the input full spectrum, which indicates that coefficients $a_i$ of the best-fitting small-scale component derived from our method can recover the input full spectrum to a well degree of accuracy.

(iv)Finally, using these coefficients $a_i$, we estimate the average age and metallicity of the best-fitting solution using the following equations:

\begin{equation}\label{meanage}
   log(Age)=\frac{\sum\limits_{i=1}^{N} a_{i}log(t_{i})}{\sum\limits_{i=1}^{N} a_{i}}
\end{equation}
and
\begin{equation}\label{meanmetal}
  [M/H]=\frac{\sum\limits_{i=1}^{N} a_{i}[M/H]_{i}}{\sum\limits_{i=1}^{N} a_{i}} ,
\end{equation}
where $t_{i}$ and $[M/H]_{i}$ represent the age and metallicity of the $i$th SSP. Here, we provide light-weighted mean stellar population parameters.
The comprehensive tests of our method are performed in next section.

\section{Testing our fitting method}

To validate our fitting algorithm, we make tests using two sets of spectra: mock galaxies and real astronomical galaxies from LAMOST and SDSS observations. The testing using mock galaxies based on SSPs is to explore possible bias in our fitting algorithm by comparing the derived properties with the intrinsic values built in the mock spectra. The testing using spectra of LAMOST and SDSS galaxies allows us to compare the stellar population properties derived from LAMOST spectra using our method with their SDSS counterparts(which are believed to have high accuracy of flux calibration) obtained from full spectral fitting.

\subsection{Testing with mock galaxies}
\label{sec:testmock}

We create mock galaxies using the SSP templates from V10 \citep{Vazdekis2010}. We choose SSPs covering 25 ages from 0.06Gyr to 15Gyr and four metallicities from -1.71 to 0.22 (log(Z/Z$\sun$)) to generate the mock spectra. For each metallicity, we randomly pick one from the 25 SSPs with different ages, which results four selected SSPs with different metallicities and ages. Then, a combination of the four SSPs with random weights can generate one mock spectrum. The intrinsic mean age and metallicity of this mock spectrum are computed by combining ages and metallicities of the four SSPs with their corresponding weights. We repeat the above step 1000 times and then build 1000 synthetic mock spectra. In addition, we create about 200 high metallicity(-0.25, 0.22) spectra to supplement the high metallicity end of mock sample. In order to better simulate the real spectra, we redden the spectra in the mock sample by Calzetti dust extinction curve \citep{Calzetti2000} assuming color excesses E(B - V) randomly selected from range(0.01,0.2).

And then, we apply a Gaussian perturbation to each flux pixel of these mock spectra with S/N = 5, 10, 20 and 30, as described by the following equation:

\begin{equation}\label{equ_mock}
  F_{mock'}(\lambda_{i})= F_{mock}( \lambda_{i}) + N \Bigg ( 0, \Bigg ( \frac{F_{mock}( \lambda_{i})}{S/N} \Bigg )^2 \Bigg) ,
\end{equation}
where $F_{mock'}(\lambda_{i})$ is the flux  of the mock galaxy at wavelength $\lambda_{i}$, $F_{mock}(\lambda_{i})$ is its corresponding flux free of noise, and $N ( \mu, \sigma^2)$ is a Gaussian perturbation with $\sigma^2$ characterized by the given S/N.

Finally, we obtain a total of 4800 mock spectra as our testing sample. For each S/N, there are 1200 mock spectra in our testing sample.

\begin{figure}
\centering
	\includegraphics[width=9cm]{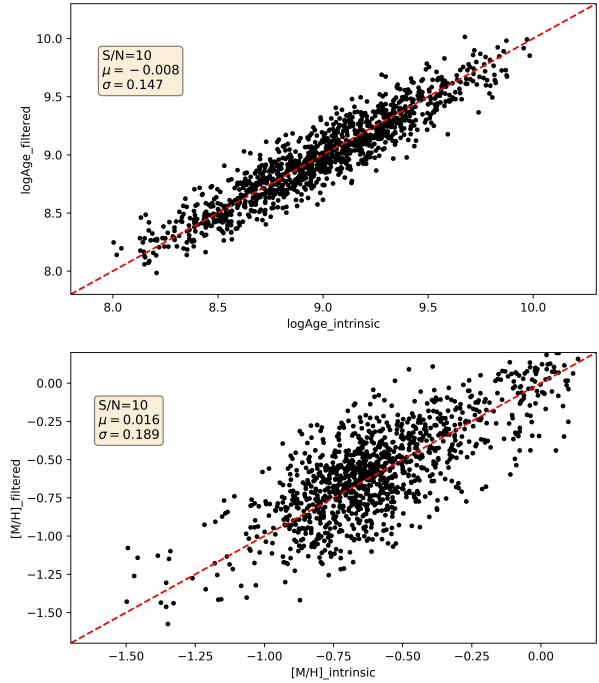}
    \caption{Comparisons of the ages and metallicities between our derived parameters and intrinsic values of the mock spectra with S/N = 10. The top panel shows the comparison of ages in log(yr) and the bottom displays the comparison of metallicities in log(Z/Z$\sun$). The mean and standard deviation of the differences are given in the top left corner of each panel. Note that the points out of 3$\sigma$ are clipped in each panel. The red line is the identity line (y = x).}
    \label{fig:compare_mock}
\end{figure}

\begin{figure}
\centering
	\includegraphics[width=9cm]{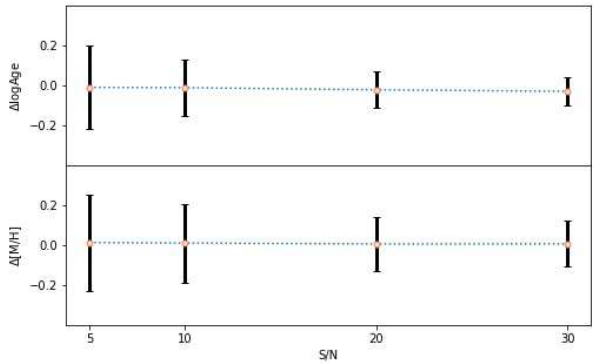}
    \caption{Differences between the stellar population of mock spectra derived by our method and their intrinsic values as a function of $S/N$. The error bars express the median values and the standard deviations of the differences in each $S/N$ bin. There are clear declines with $S/N$ increasing.}
    \label{fig:mock_para_sn}
\end{figure}

We run the fitting process detailed in Section \ref{sec:fittingM}, and measure the mean age and metallicity of each mock spectrum. We then compare the fitting results with the intrinsic values of mock spectra, which are illustrated in Figure \ref{fig:compare_mock} and Figure \ref{fig:mock_para_sn}. Figure \ref{fig:compare_mock} shows the comparison of the fitted mean ages and metallicities with intrinsic values for the mock spectra with S/N=10, where good linear correlations are clearly seen. The mean($\mu$) and standard deviation($\sigma$) of differences between our fitting results and the intrinsic values are also indicated in each panel. For age, the standard deviation is $\sim$ 0.15. For metallicity, the standard deviation is slightly larger with the value $\sim$ 0.19. In Figure \ref{fig:mock_para_sn}, we describe the variation of the differences between our derived parameters and the intrinsic values along with S/N. As expected, the deviation decreases when S/N increases. For stellar age, the consistence is within 0.2 dex even down to S/N=5. The deviation of the stellar metallicity measurement is large, which may be caused by the less metallicity grids used in fitting or the complexity of the metallicity measurement from the galaxy spectrum\citep{Girardi2000, Bruzual2003, Gallazzi2005}.

As we can see from the testing results of mock galaxy spectra that our method based on small-scale component works well even if we consider dust attenuation effect into mock spectra. In fact, the dust attenuation effect of spectrum as the large-scale component is filtered out during our fitting process.

\subsection{Testing with SDSS galaxies}
\label{sec:comp_SED}

In this section, we further test our fitting code using the spectra of real galaxies. We make comparisons of the fitting results on the galaxy spectra with both observations in SDSS and LAMOST spectra survey. In specific, we derive the stellar population parameters from the LAMOST spectra using the method outlined above, and compare with results from SDSS spectra using the classical full spectral fitting.

We first get the same target observations from LAMOST and SDSS by cross-matching the galaxy spectra in LAMOST DR7 with the counterparts in SDSS DR16 with $S/N_{r}\geq5$. In order to eliminate the difference of S/N between LAMOST and SDSS spectra, we restrict the counterparts in two samples with $|S/N_{r}\_LAMOST - S/N_{r}\_SDSS|<1$. We then obtain $\sim$7000 spectra of LAMOST and their SDSS counterparts as our testing samples. For LAMOST spectra, we use our method to derive the stellar population parameters based on the small-scale features, while for SDSS spectra we use pPXF\citep{Cappellari2004, Cappellari2017} to make the full spectral fitting. The templates we used in pPXF are the same 36 MILES SSPs as our method described in Section \ref{sec:fittingM}. During the pPXF fitting, we adopt a high-order multiplicative polynomial to mimic the spectra reddening effect and without adopting a specific reddening curve. In addition, we don't apply a regularisation to the star formation history in pPXF fitting, because we focus on the weighted mean age and metallicity of galaxy rather than smooth weights.

\begin{figure*}
\centering
	\includegraphics[width=16cm]{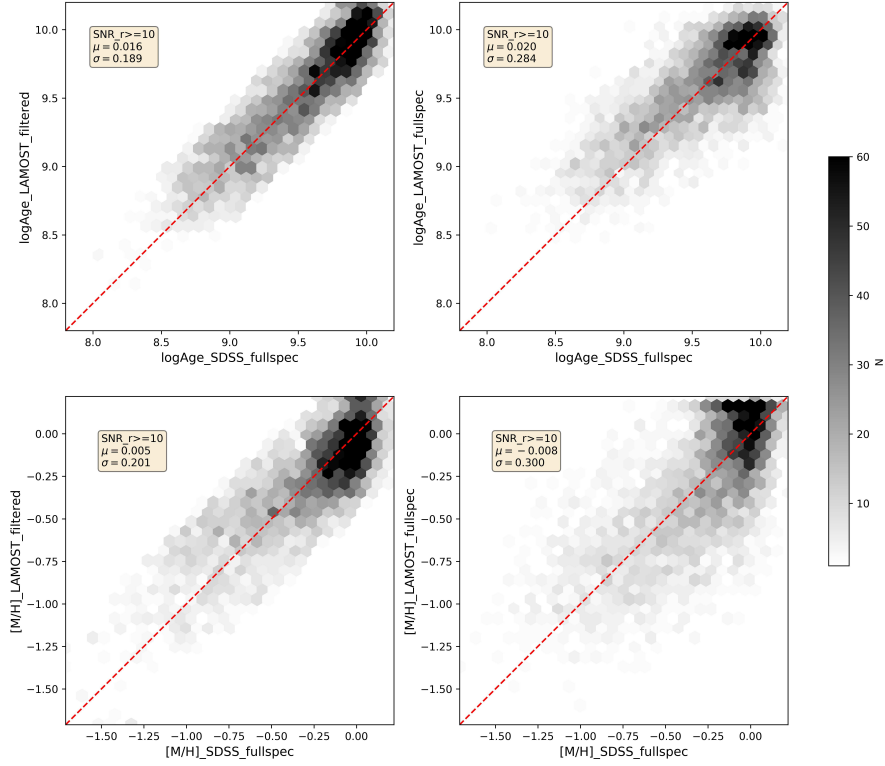}
    \caption{Comparisons of the age and metallicity measurements for the galaxies with both LAMOST and SDSS spectra( $S/N_{r}\geq10$).
    The left two panels show the comparisons between the LAMOST spectra fit with our small-scale feature method and their SDSS counterparts fit with the full spectral fitting. While the right two panels compare the results from the same full spectrum fitting method for LAMOST and SDSS spectra. Note that the points out of 3$\sigma$ are clipped in each panel. The red line is the identity line (y = x).}
    \label{fig:compare_LAMOST}
\end{figure*}

The comparisons of the stellar population parameters for LAMOST galaxies and their SDSS counterparts with $S/N_{r}\geq10$ are shown in the left two panels of Figure \ref{fig:compare_LAMOST}. As can be seen, the agreements are very well, where the mean differences are only 0.016 and 0.005 dex and the 1$\sigma$ uncertainties are 0.189 and 0.2 dex for age and metallicity respectively. The results prove again the validity of our method, that is, we can recover the stellar population properties of galaxies well based on small-scale features only.

We use pPXF to make the same full spectrum fitting on the LAMOST spectra in our testing sample, and compare the derived LAMOST parameters with their SDSS ones. In this case, the templates and other settings of pPXF are the same as the above. And pPXF also employs high-order multiplicative polynomials. The comparison of the fitting results on the LAMOST and SDSS spectra using the standard full spectrum fitting is shown in the right panel of Figure \ref{fig:compare_LAMOST}. As can be seen, the consistence now is much worse than the left two panels. This result illustrates the importance of the accuracy of the flux calibration in the full spectrum fitting. This is precisely the reason for the need to develop the small-scale feature fitting method for LAMOST spectra in this paper.

\begin{figure}
\centering
	\includegraphics[width=9cm]{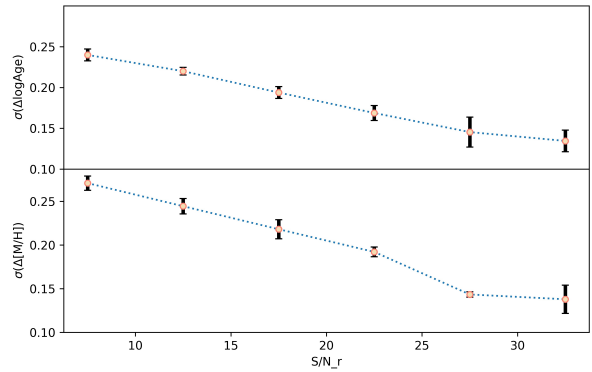}
    \caption{Dispersion of the differences of the stellar population parameters between the LAMOST spectra measured by our method and their SDSS spectral counterparts using the full spectral fitting as a function of $S/N_{r}$. The top panel shows the dispersion of $\Delta logAge$ and the bottom panel plots that of $\Delta [M/H]$. The error bar expresses the median value and the standard deviation of the differences in each $S/N_{r}$ bin. There are clear declines with $S/N_{r}$ increasing.}
    \label{fig:para_sn}
\end{figure}

In addition, we consider the effect of $S/N_{r}$ on the difference of the stellar population  parameters between our method on LAMOST spectra and pPXF full spectrum fitting on SDSS spectra. Figure \ref{fig:para_sn} displays the dispersion of the differences as a function of $S/N_{r}$. We see that as the $S/N_{r}$ increases, the dispersion of the differences decreases significantly. From these comparisons, we see that our small-scale feature fitting method can obtain reliable estimation(with an accuracy better than 0.25 and 0.3 dex for age and metallicity respectively) on the average stellar populations of galaxies for the LAMOST spectra with $S/N$ down to 5.

In conclusion, as illustrated by the above tests, our small-scale feature fitting method is fully adapted to the estimation of the stellar population parameters for the LAMOST galaxy spectra. The main benefit of our method is to avoid possible biases from the uncertain flux calibration of the LAMOST spectroscopy.

\section{Application to LAMOST galaxy spectra}
\label{sec:apptoLAMOST}

We apply our small-scale feature fitting algorithm to the LAMOST galaxy spectra that have not been spectroscopically targeted by SDSS to derive their stellar population parameters. After fitting, with an extensive visual inspection, we exclude a few spectra( less than 1.5 percent) which are badly fitted. Finally, we have obtained the stellar population parameters of $\sim$ 43,000 galaxies( If a galaxy with more than one spectrum, we keep the spectrum with highest $S/N_{r}$).

\subsection{The catalogue of stellar population properties for LAMOST galaxies}

We present the catalogue of our derived stellar population properties( age and metallicity) as a value added catalogue for $\sim$ 43,000 galaxies in LAMOST DR7. All galaxies in our catalogue are only spectroscopically observed by LAMOST but without spectral counterparts in SDSS DR16. Table \ref{tab:catalog} lists a part of this catalogue. The complete catalogue is available online.

\begin{table*}
\centering
   \renewcommand\arraystretch{1.4}
	    \caption{Catalogue of the stellar population parameters for $\sim$ 43,000 galaxies in LAMOST DR7.}
	    \begin{tabular}{p{1.4cm}p{1.3cm}p{1.3cm}p{1.1cm}p{0.7cm}p{0.7cm}p{0.7cm}p{1.0cm}}
		\hline
		obsid & ra & dec & z & $S/N_{r}$  & $mag_{r}$ & age & [M/H]\\
		(1) & (2) & (3) & (4) & (5) & (6) & (7) & (8)\\
		\hline
        115042    & 332.08942 & 1.57701   & 0.076754 & 6.48  & 18.071 & 8.92 & -1.141 \\
        173403192 & 20.888573 & -2.45587  & 0.045233 & 27.39 & 15.471 & 10.119 & 0.024\\
        173503204 & 344.96793 & 7.02769   & 0.041176 & 26.52 & 15.961 & 10.142 & 0.011\\
        266716179 & 2.86529   & 3.20193   & 0.153985 & 13.38 & 17.661 & 9.407 & -0.146\\
        369901214 & 0.01203   & 5.59878   & 0.039678 & 10.11 & 16.869 & 9.945 & -0.456\\
        369904055 & 0.338594  & 8.157074  & 0.038387 & 43.5  & 15.332 & 9.873 & -0.219\\
        399414225 & 9.502266  & 3.7608092 & 0.129217 & 10.65 & 16.080 & 9.618 & -0.118\\
        486912170 & 28.841459 & 33.876765 & 0.223634 & 9.28  & 18.280 & 9.223 & -0.626\\
        398109210 & 28.861072 & 2.8273228 & 0.182454 & 12.42 & 17.233 & 8.873 & -0.647\\
        723705095 & 136.57802 & 9.195647  & 0.045942 & 22.64 & 16.374 & 9.273 & -0.549\\

		\hline
	\end{tabular}
	\begin{tablenotes}
        \item[1]Notes:Columns (1)--(5) are retrieved from the LAMOST published catalogue: obsid -- unique spectra ID of LAMOST, ra -- right ascension (J2000), dec -- right ascension (J2000), z -- redshift, and $S/N_{r} - $signal to noise ratio in $r$ band;

            Columns (6) is Petrosian magnitude in $r$ band cross-matched with the photometric catalogue of SDSS;

            Columns (7) is the mean stellar age in log(yr);

            Column (8) is the mean  stellar metallicity in log(Z/Z$\sun$).
      \end{tablenotes}
\label{tab:catalog}
\end{table*}

\begin{figure}
	\centering
	\includegraphics[width=9cm]{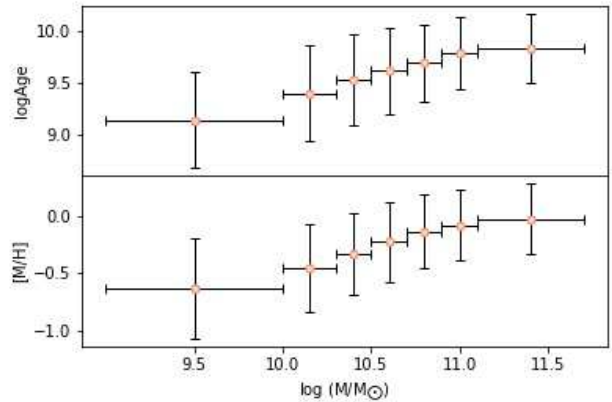}
    \caption{Distributions of the age and metallicity of galaxies in our catalogue as a function of stellar mass(in solar units). The top panel shows the age distribution in log(yr), whereas the bottom panel shows the distributions of metallicity in dex. The error bars show the 1$\sigma$ in the age and metallicity (vertical error bars) and the stellar mass range used for each bin (horizontal error bars).}
    \label{fig:LAMOST_para_mass}
   \end{figure}

Figure \ref{fig:LAMOST_para_mass} shows the distribution of age and metallicity in our catalogue for different bins of stellar mass. The stellar masses of our galaxies are estimated using magnitudes in $r$ band and stellar mass-to-light (M/L) ratios. The M/L is derived from a function of $g - r$ color given by \citet{Bell2003}. We see from Figure \ref{fig:LAMOST_para_mass} that the age and metallicity of galaxies increase as stellar mass increases.

\subsection{An scientific application of the stellar population properties of LAMOST spectra}

An interesting science target selection in the LAMOST spectral survey is the complementary galaxy sample which is a subset of the main galaxy sample(MGS) of SDSS without spectroscopic observations till DR7. We remind that the MGS of SDSS is a magnitude complete galaxy sample down to $<17.77$\citep{Stoughton2002}. As shown by \citet{Shen2016}, those MGS in SDSS without spectroscopic observations are mainly caused by the fiber collision effect, so these galaxies targeted in LAMOST are probably in galaxy pair environment. As a result, some galaxies in our catalogue are indeed members of galaxy pairs.

Here we present an application of the stellar population measurements of the galaxies in our catalogue, the Holmberg effects of the galaxies in pairs, which tells that the physical properties of the pair members are correlated(e.g. color)\citep{Holmberg1958, Tomov1978, Allam2004, Cao2016}. However, whether this correlation is caused by a nature or nurture effect is still not clear\citep{Allam2004, Deng2010, Melnyk2012}.

Following \citet{Feng2019},  we select the galaxy pairs using two criteria about the line-of-sight velocity difference($\Delta V$) and the projected distance($r_{p}$) : $\Delta V \leq $ 500 km s$^{-1}$ and 10 $h_{100}^{-1}$ kpc $\leq$ $r_{p}$ $\leq$ 200 $h_{100}^{-1}$ kpc. By matching galaxies in our catalogue with ones in SDSS MGS by the above criteria, we obtain a sample of $\sim$ 3,000 LAMOST-SDSS galaxy pairs. To better outline the Holmberg effect, we construct a control sample to our $\sim$ 3,000 LAMOST members in pair sample by the process similar to \citet{Ellison2008}. To be specific, a control sample is compiled by matching each pair galaxy in the same mass and redshift distributions with the galaxies with no close companions in SDSS MGS. We confirmed that the galaxies in the control sample are physically un-correlated ($r_p> 200 h_{100}^{-1}$ kpc), and also have nearly identical distributions of redshift and stellar mass to the pair sample.

\begin{figure}
\centering
	\includegraphics[width=8.5cm]{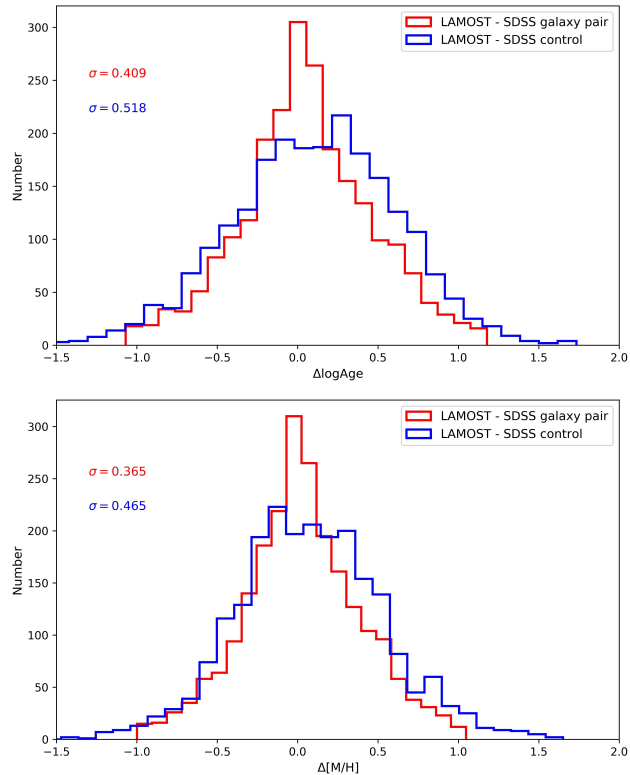}
    \caption{Distributions of the differences of age and metallicity of two members in the pair sample and control sample. In each panel, the red histograms express the differences of two companions in pair sample, and the blue histograms describe the differences of the control galaxies. The standard deviations ($\sigma$) of $\Delta$logAge and $\Delta$[M/H] of two sample are labeled in the top left of each panel with the same color coding as the histograms.}
    \label{fig:pairs_para}
\end{figure}

The differences of the ages and metallicities of two members of the pair and control sample are shown in Figure \ref{fig:pairs_para}. In each panel, the red histograms represent the differences of members in the pair sample, while the blue histograms show the differences in the control sample. The standard deviation ($\sigma$) of $\Delta$logAge and $\Delta$[M/H] are indicated in the red and blue with the same color meaning as histograms. From the figure, we see that the distributions of differences of the age and metallicity of the paired galaxies are both have significantly smaller scatter than the control galaxies. Since the pairs and their controls have the same mass and redshift, the correlation between the magnitudes (so that colors) of the pair members has been reflected in the control sample. If we consider the correlation between the masses of pair members as a nature effect \citep{Feng2019}, the smaller scatter of $\Delta log Age$ and $\Delta[M/H]$ we see for the pair members should originate from co-evolution of pair members, i.e. the nurture effect.

\section{Summary}

In this study, we explore a fitting method based on the small-scale features of galaxy spectra to derive the basic stellar population parameters, the mean stellar age and metallicity, for the LAMOST galaxy spectra. The core of our method is to separate the small-scale features from the large-scale components and only fit the small-scale features of the observed spectra to the small-scale features of models using Equation \ref{eq:chi2}. Generally, the large-scale component is regarded as the continuum shape, while the small-scale component represents the spectral lines. So our method mainly relies on absorption line features in galaxy spectra independently of the continuum shape. One important advantage of this method is that it is suitable for the stellar population analysis of galaxies with uncertain flux calibration or spatially non-uniform dust attenuation. We check the performances of our method using mock galaxies and real observed spectra of LAMOST and SDSS galaxies, and the results show a well consistency within 0.2 dex despite higher uncertainties(0.25 dex and 0.3 dex for age and metallicity, respectively) for galaxies in S/N=5.

Another goal of this paper is to present a public catalogue of stellar age and metallicity for galaxies only spectroscopically observed in LAMOST DR7 but without spectral counterparts in SDSS DR16 using our fitting method. As a result, a catalogue of stellar population parameters for $\sim$43000 galaxies is published, which is the first time to estimate the age and metallicity for galaxies in LAMOST spectral survey. We give a preliminary study on the Holmberg effect of galaxy pairs in our catalogue, which may play an important role in the study of the physical properties of the low redshift galaxies and galaxy systems. In the next work, we plan to apply our small-scale feature fitting to all LAMOST galaxies, and continuously update the stellar population catalogue as a value added catalogue available on LAMOST official website.

\acknowledgments

This work was supported by the National Natural Science Foundation of China (grant Nos. 11903008, U1931106, 12073059), National Key R\&D Program of China No. 2019YFA040550, Shandong Provincial Natural Science Foundation of China (grant No. ZR2019YQ03), Shandong QingChuang science and technology plan (grant No. 2019KJJ006) and the Science Foundation of Dezhou University(grant No.2019xjrc39).

The Guoshoujing Telescope (the Large Sky Area Multi-Object Fiber Spectroscopic Telescope LAMOST) is a National Major Scientific Project built by the Chinese Academy of Sciences. Funding for the project has been provided by the National Development and Reform Commission. LAMOST is operated and managed by the National Astronomical Observatories, Chinese Academy of Sciences.

%




\appendix

\section{Modeling the galaxy spectrum based on small-scale component} \label{sec:model_decomp}

In the conventional full spectral fitting, the spectrum of a galaxy $S(\lambda)$ could be modelled by a linear combination of SSPs,

\begin{equation}\label{eq:appendix_full}
   S(\lambda)=\Bigg [\sum\limits_{i=1}^{N} a_{i}SSP_{i}(\lambda)\Bigg ] \cdot \mathcal{D}(\lambda)
   \,,
\end{equation}
where  $a_i$ is the fraction of the  stellar population $SSP_i$,  $\mathcal{D}(\lambda)$ represents modification of the spectrum by either the uncertainties of flux calibration or the  dust attenuation effect.

Following \citet{Li2020}, we define the large-scale component of a galaxy spectrum  as,

\begin{equation}\label{eq:appendix_large_simple}
   S_{L}(\lambda)=\frac{1}{\Delta\lambda}\int_{\lambda-\Delta\lambda/2}^{\lambda+\Delta\lambda/2} S(\lambda')d\lambda'\\
   \,,
\end{equation}
where $\Delta\lambda$ is the wavelength window size. Using Equation \ref{eq:appendix_full} into the above equation, we  get

 \begin{equation}\label{eq:appendix_large}
\begin{split}
   S_{L}(\lambda)&=\frac{1}{\Delta\lambda}\int_{\lambda-\Delta\lambda/2}^{\lambda+\Delta\lambda/2} \Bigg [\sum\limits_{i=1}^{N} a_{i}SSP^i(\lambda')\Bigg ]\cdot \mathcal{D}(\lambda') d\lambda'\\
   &\approx\frac{\mathcal{D}(\lambda)}{\Delta\lambda}\int_{\lambda-\Delta\lambda/2}^{\lambda+\Delta\lambda/2} \Bigg [\sum\limits_{i=1}^{N} a_{i}SSP^i(\lambda')\Bigg ] d\lambda'\\
   &=\Bigg [\sum\limits_{i=1}^{N} a_{i}SSP_L^i(\lambda)\Bigg ]\cdot \mathcal{D}(\lambda)
   \,,
\end{split}
\end{equation}
where $SSP_L^i$ is the large-scale component of  $SSP_{i}$ following the same smooth process of Equation \ref{eq:appendix_large_simple}. The approximation in this equation is based on  the fact that $\mathcal{D}(\lambda)$ is a large-scale correction factor and could be approximated as a constant inside the window function $\Delta\lambda$. Then, the  small-scale component of the galaxy spectrum is simply written as

\begin{equation}\label{eq:appendix_small}
   S_{S}(\lambda)= \frac{S(\lambda)}{S_{L}(\lambda)}= \frac{\sum\limits_{i=1}^{N} a_{i}SSP_{i}(\lambda)}{\sum\limits_{i=1}^{N} a_{i}SSP_L^i(\lambda)}
   \,.
\end{equation}

If the stellar populations of the galaxy we study can be well approximated by one SSP, we have
\begin{equation}
   S_{S}(\lambda)= \frac{SSP_i(\lambda)}{SSP_{L}^i(\lambda)}\equiv SSP_S^i(\lambda)
   \,,
\end{equation}
where $SSP_{S}^i$ is defined as the small-scale component of SSP accordingly. That is to say, in this case($N=1$), we can model the small-scale component of the galaxy with $SSP_{S}^i$ directly and so that eliminate the unknown factor $D(\lambda$ ). On the other hand, when the stellar populations of the galaxy are complex($N>1$), we can not fit $S_S(\lambda)$ with a linear combination of $SSP_S^i$, because
\begin{equation}
   S_{S}(\lambda)= \frac{\sum\limits_{i=1}^{N} a_{i}SSP_{i}(\lambda)}{\sum\limits_{i=1}^{N} a_{i}SSP_L^i(\lambda)} \neq \sum\limits_{i=1}^{N} a_{i}SSP_S^i
   \,.
\end{equation}

\section{The choice of window size for calculating the large-scale component} \label{sec:winsize}

\begin{figure}
	\centering
	\includegraphics[width=9cm]{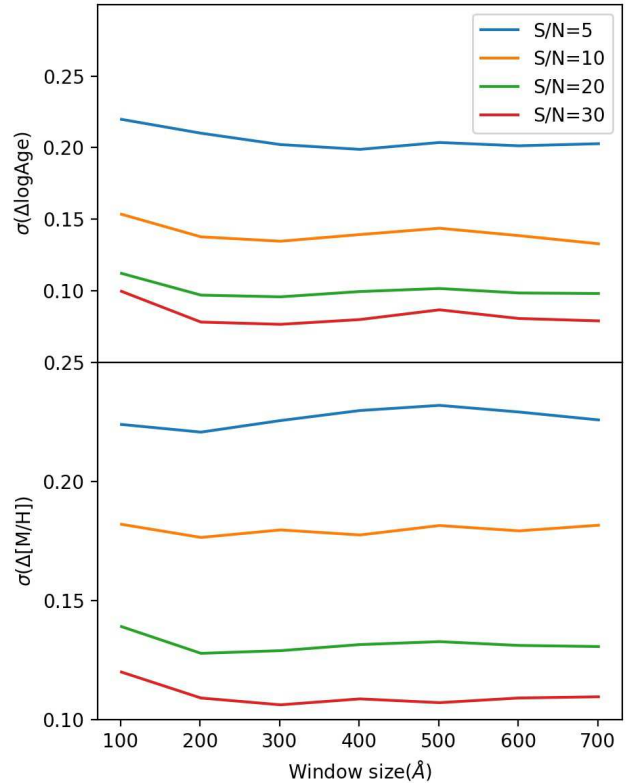}
    \caption{Testing on window sizes for calculating the large-scale component in our fitting method using the mock galaxies created in Section \ref{sec:testmock}. The standard deviation $\sigma$ of the difference between our derived stellar populations(age and metallicity) and intrinsic values is displayed as a function of the window sizes from 100 to 700\AA \ for four different S/Ns(5, 10, 20, 30).}
    \label{fig:test_winsize}
   \end{figure}

We use average filtering to calculate the large-scale component of spectrum. The sliding window size is an important factor in the calculation. In order to find a suitable window size, we test different window sizes ranging from 100 to 700\AA \ in our fitting method using the mock galaxies created in Section \ref{sec:testmock}, and compare the standard deviation $\sigma$ of the difference between our derived stellar populations and intrinsic values. Figure \ref{fig:test_winsize} plots $\sigma$ as a function of the window sizes for four different S/Ns(5, 10, 20, 30). We see that the changes of $\sigma$ do not have effects on our fitting results when window size $>$ 200\AA\ . Generally, the window size is larger, the computational time is larger. In order to balance the window size and computational time, we finally choose 300\AA \ as the window size for calculating the large-scale component.

\bibliography{sample63}{}
\bibliographystyle{aasjournal}



\end{document}